\documentstyle[floats,aps]{revtex}
\begin{document}
\draft
\preprint{}
\twocolumn[\hsize\textwidth\columnwidth\hsize\csname @twocolumnfalse\endcsname
\title{
Liquid Crystal Analogue of Abrikosov Vortex Flow in Superconductors}
\author{A.Tanaka,${}^{1,2}$ T. Ota,${}^{2}$ and R. Hayakawa${}^{2}$ }
\address{
Solid State Laboratory, Tokai Establishment, Japan Atomic Energy Research Institute, Ibaraki, Japan ${}^{1}$ \\
Department of Applied Physics, The University of Tokyo, 7-3-1 Hongo, Bunkyo-ku, Tokyo 113, Japan ${}^{2}$  }

\date{}

\maketitle
\begin{abstract}
We extend the correspondence between the Renn-Lubensky Twist-Grain-Boundary-A
 phase in chiral liquid crystals and the Abrikosov mixed state in 
superconductors to dynamical aspects. 
We find that for a TGB sample with free boundaries, an external
electric field applied along the helical axis induces a uniform
translational motion of the grain boundary system - an analogue 
of the well-known mixed state flux flow. Likewise, an analogue 
of the mixed state Nernst effect is found. In much the same way in 
which the flux flow carries intercore electric fields generating 
Joule heat in an otherwise dissipation-free system, the 
grain boundary flow carries along polarized charges, resulting 
in a finite electric conductivity in a ferroelectric.   
\end{abstract}

\vskip2pc]


The formal analogy of the smectic free energy within a Ginzburg-Landau 
approach, with that for the well known case of a superconductor 
has had a long history in the study of liquid crystals. 
It was first pointed out by de Gennes that the essential 
mean field features of 
the phase transition between the Nematic (N) and Smectic-A (SmA)
phases  
can be derived by simply resorting to the correspondence with the 
Normal-Superconductor transition.\cite{dege} Corrections to this
mean-field picture was studied by means of a large-N expansion of 
the GL energy functional, in which a runaway RG trajectory was 
interpreted as an indication of a fluctuation-driven first order 
transition.\cite{hlm} 
\ \par

An important application of this analogy was carried out by 
Renn and Lubensky, who, based on the mean-field phase diagram 
of type-II superconductors in an applied magnetic field,
 predicted an intermediate phase 
lying between the cholesteric (Ch) and SmA phases.\cite{rl}  
This new phase, termed the Twist-Grain-Boundary-A(TGBA) phase, is the 
counterpart of the Abrikosov vortex lattice phase (also 
known as the mixed phase), and is composed of a regular 
network of screw dislocations. For the detailed structure of this 
lattice, readers are reffered to ref.\cite{rl}. 
The TGBA phase was also independently discovered by
exerimenters\cite{good}, and immediately became a 
subject of intense research. \ \par
So far, theoretical investigations on the TGBA based 
on the superconductor analogy have been restricted to mean-field, 
static aspects. (See however refs\cite{toner,toki,kam}.)  
Considering however the rich variety of 
mixed state physics that continues to be revealed\cite{los}
through recent (high Tc-motivated) studies of 
fluctuation effects and vortex dynamics, 
it is natural to ask what new feature arises when one 
seeks to extend the analogy in similar directions. 
As in high Tc superconductors, it is expected that 
thermal fluctuations have profound effects on the 
TGBA lattice structure; a simple estimate via the 
Ginzburg criteria strongly suggests this vulnerability, and 
there is an additional intrinsic instability towards disorder 
which is a consequence of the very symmetry of the TGB.\cite{toner,ota}  
However, a detailed study of the full effect of thermal fluctuation 
within the GL formalism requires considerable computing efforts, 
and we choose to set a less ambitious goal. 
In this paper, we study the fundamental properties of dislocation motion 
in the TGB system. 
As a result, we obtain several new entries to the list of 
superconductor-liquid crystal analogue, as sumarised in Table 1.
A major feature is the precise liquid crystal 
analogue of the
flux flow motion of the Abrikisov vortex lattice. 
Although the long wavelength hydrodynamics of the TGBA phase has
been previously worked out,\cite{hat} this paper is, to our
knowledge, the first one which has considered the grain boundary
itself and its constituent screw dislocation lines as mobile entities.
An important implication of this motion is
that the dislocation cores carry electric charges along with the flow 
through the ferroelectric media. Thus the system is predicted to be
conductive. 
\ \par 

To generate the motion of the dislocations, we will take 
advantage of the chiral nature of system. 
Due to the lack of chiral symmetry, 
polar vector fields can couple linearly to axial vector fields.  
If we consider the torque of the smectic layer normal as the latter  
and couple it to an external polar vector field, this will effectively
exert a force on the grain boundary. As the external field, we
consider  
(1) an electric field applied along the helical axis of TGBA phase 
and (2) temperature gradient imposed along the same direction. 


In the SmA phase of such materials, it is well known that when an electric field is applied in-plane to the smectic layers, a finite tilting of the molecular axis from the smectic layer normal is favored. 
This phenomenon
is called the electroclinic effect.\cite{mey}    
We shall show here that the stress induced by the electroclinic effect produces the driving force acting on screw dislocations. 
Our starting point is the following free energy, which is a 
combination of the  covariant  Chen-Lubensky(CL) model\cite{rl} and the energy related to the electroclinic effect.\cite{mey}: 
\begin{eqnarray}
\lefteqn{
F=\int d^3x [  r \mid \psi \mid ^2 +C\mid (\nabla -iq_0 {\bf n})\psi \mid ^2+{g\over 2}  \mid \psi \mid ^4  } \cr
& &+ {{K_1}\over{2}} (\nabla \cdot {\bf n})^2 +{{K_2}\over{2}} ({\bf n} \cdot \nabla \times {\bf n} -k_0 )^2 +{{K_3}\over{2}}({\bf n} \times
\nabla \times {\bf n})^2 \cr 
& &+{ {{{\epsilon}_0}E^2}\over{2} }+ {{P^2}\over{2\chi}}- \eta ({\bf n}_0 \times (\nabla u-\delta {\bf n}))\cdot {\bf P}], \label{fe} 
\end{eqnarray}
where, ${\bf n}$ is the Frank director, $\psi ({\bf x})$ is the
smectic order parameter, 
$\psi ({\bf x})=\mid {\psi} \mid e^{iq_0 u} e^{{iq_0 {\bf n}_0} \cdot {\bf x}}$ 
with $q_0=2 \pi /d$, $d$ the separation between smectic layers, 
$u$ the smectic layer displacement, ${\bf n}_0$ is the normal vector
of smectic layer.  
$\eta$ is the coupling constant between the polarization ${\bf P}$ 
and ${\bf n}_0 \times (\nabla u-\delta {\bf n})$, and $\chi$ is the electric susceptibility. 
The terms 
$\eta ({\bf n}_0 \times (\nabla u-\delta {\bf n}))\cdot {\bf P} $ 
and $K_2k_0 {\bf n}\cdot \nabla \times {\bf n}$ 
reflect the mentioned symmetry of the system.
Differentiating $F$
with respect to $\delta {\bf n}$, 
gives the stress field 
${\bf J}\equiv - \frac{\delta F'}{ \delta \delta {\bf n}} 
=B' ({\bf v} -\delta {\bf n})$ 
with ${\bf v}=\nabla u$, 
and $B'=2Cq_0^2{\mid \psi \mid }^2-{\eta}^2 \chi$.
${J}$ is analogous to the 
supercurrent ${\bf J}_s$ which is 
the free energy differentiated with respect to the vector potential ${\bf A}$.
$x$-axis, 
>From the solutions of $\nabla \cdot {\bf J}=0$ and $ {{\delta
F}\over{\delta {\bf n}}}=0$, 
we see that ${\bf v}-\delta {\bf n}_v-\delta {\bf n}_{e}$ 
is divided into two parts. 
One is the ^^ ^^ vortex'' part ${\bf v}-\delta {\bf n}_v $,\cite{day}
with nonvanishing curl, 
and the other is the uniform part,
$\delta {\bf n}_{e} = 
{{ \eta \chi ({\bf n}_0 \times {\bf E})}\over {B'}}$,
induced by the electric field ${\bf E}$.
Applying the virtual work principle, we find 
a Peach-Koehler tyoe force 
\begin{equation}
{\bf f}_l= {\bf J}_e \times d\hat{{\bf z}}. \label{vir3}
\end{equation}
acting on each unit length of the dislocation. 
Here $d \hat{\bf z}$ is the Burgers vector. 

Now we compare this with superconductors. 
$B'({\bf v} -\delta {\bf n}_v)$
corresponds to the vortex part of the supercurrent, and  
$-B'\delta {\bf n}_e$ corresponds to the transport current. 
In the superconductor case, the corresponding force 
${\bf f}_L={\bf J}_t\times{\Phi}_0 \hat{{\bf z}}$ 
is just the Lorentz force acting on the vortex line per unit length. 
${\Phi}_0$ is the flux quantum.
Comparing ${\bf f}_l$ with ${\bf f}_L$, 
it is clear that the force acting on the single screw dislocation
corresponds to Lorentz force in superconductors.

Next we consider the case where the Peach-Koehler force acts 
on the grain boundaries which is just an array of screw dislocations. 
For the grain boundary located at $x=0$, 
the unit normals ${\bf n}_0^L$ and 
${\bf n}_0^R$ of the SmA slab at $ x <0$ and $x > 0$ , 
respectively are forced to be rotated with respect to 
each other around the $x$-axis by an angle of $2\pi \alpha$. 
A uniform electric field ${\bf E}=E\hat{{\bf x}}$ applied along the
$x$-axis will generate the stress ^^ ^^ currents'' on the left and right ends
of each SmA slab given by
${\bf J}_e^L=-\eta \chi {\bf n}_0^L\times {\bf E} \label{jl}$, 
and ${\bf J}_e^R=-\eta \chi {\bf n}_0^R\times {\bf E}$, 
respectively (See Fig1). 
The extension of eq.(\ref{vir3}) 
to this configuration is straightforward and 
the force ${\bf f}_{tgb}$ acting on the grain boundary 
per unit area is 
\begin{equation}
{\bf f}_{tgb}=
-{{\eta \chi Ed}\over{l_d}} \cos(\pi \alpha)\hat{{\bf x}} 
\sim -2\eta \chi E \pi \alpha \hat{{\bf x}}, \label{fgt}
\end{equation}
where the final expression is valid for small $\alpha$. 
This simple superposition is invalidated near $k=k_{c2}$ where 
dislocations cores begin to overlap. 


Next we consider the case where a temperature gradient is imposed along the pitch. 
In analogy with the Nernst effect in a superconducting mixed 
phase,  
it is anticipated that screw dislocations in the TGBA phase a+re 
influenced by a thermal force generated in the direction of the temperature gradient. 
We shall show below that this is indeed the case.
The physics of what is shown below may be summarized as follows. 
The temperature gradient produces the gradient of smectic density $\mid \psi \mid $, where $n_s={\mid \psi \mid}^2 $.  
This in turn disturbs the equilibrium of screw dislocations, giving
rise to an effective driving force acting on the system. 
\ \par
Following the approach of Koklov and co-workers\cite{kuk} 
who evaluated the Nernst coefficient for the superconductor case
solely within the Ginzburg-Landau formalism, 
we can derive \cite{ota} an expression for the total force acting on the ensemble of screw dislocations in the case where the temperature gradient produces a steady nonequilibrium distribution of screw dislocations. 
For the condition ({\it i.e.} ${{\xi}_{s }}^2/l_bl_d \ll 1$ with $\xi_{s}$ the smectic
correlation length),  
the effective force per unit area on the  grain boundary is 
\begin{eqnarray}
{\bf f}_x&=& 
{\Lambda} {{{\nabla}_x (\Delta n_s)}\over{n_s}} \cr
&\sim &  {{{\nabla}_x (\Delta n_s)}\over{n_s}} \left(
\frac{Bd^2}{2\pi l_d} \ln {\left( \frac{l_d}{{\xi}_s} \right) }
 +\frac{Bd^2{\lambda }_2}{4l_d^2} \exp { - \frac{l_d}{{\lambda }_2}  } \right).
\label{fns}
\end{eqnarray}
$B=2Cq_0^2{\mid \psi \mid }^2$
, and the penetration length ${\lambda}_2=\sqrt{K_2/B}$. 
It should be noted that the drift force 
{\bf f} is proportional to the gradient
of smectic density $\nabla (\Delta n_s)$, 
The first term of the right hand side of eq.~(\ref{fns}) 
has simple form of a single dislocation contribution multiplied by
a factor of $1/{l_d}$, the density of screw dislocations in a layer.
The second term reflects the global twisted structure of the TGBA system.


Now that we have obtained the forces induced by the coupling to
external fields, we now turn to the actual motion of the dislocations.

We start with the ^^ ^^ model A'' type equation for the order parameter
\cite{ho}: 
\begin{equation}
{\gamma}_{\psi}  {\partial }_t  \psi
 =- \frac{\delta F}{\delta {\psi}_{\ast}}
+ \zeta (t).
\label{ptd}
\end{equation}
$\zeta (t)$ is a thermal noise obeying the fluctuation-dissipation relation, 
$<\zeta (t) \zeta (t') >=\frac{2k_BT}{\mid {\gamma}_{\psi}\mid}
\delta(t-t')$.  
${\gamma}_{\psi}={\gamma}_{\psi}'+i{\gamma}_{\psi}''$ is generically a
complex kinetic coefficient.  
Once the defect moves, 
we will have to consider the effect of a hydrodynamic lift force
(Magnus force). 
The existence of such a force and its relevance to the Hall sign anomaly is an issue of present debate in superconductors.\cite{ao,gai} 
It can be shown that a imaginary part of the relaxation time 
${\gamma}_{\psi}''$ in eq.~(ref{ptd}) can give rise to such a force. 
However, renormarization-group calculation shows 
$({\gamma}_{\psi}''/{\gamma}_{\psi}')$ term to be an 
irrelevant perturbation in the vicinity of the transition point, {\it i. e.} near $k=k_{c2}$.\cite{dor} 
Furthermore, it was concluded in Ref.~\cite{plei} that 
the force is absent in the single screw dislocation case. 
Thus we consider it plausible to neglect such forces. 
and put ${\gamma}_{\psi}''/{\gamma}_{\psi}' = 0$. 
In this purely dissipative case, the  motion of screw dislocation 
takes the form 
${\gamma }_l{\bf v}_l={\bf f}_l$,       
where ${\gamma }_l$ is the viscosity coefficient 
for a screw dislocation and  ${\bf v}_l$ is its velocity.
$\gamma_{l}$ is determined using the  
Rayleigh functional for eq.(\ref{ptd}) 
$R[{\partial}_t\psi,{\partial}_t\psi^{\ast}] 
=\int d^3x {\gamma }_{\psi } \mid {\partial}_t\psi \mid ^2 $, 
and the balance if force 
$\frac{\partial R}{\partial {\bf v}_l}={\bf f}_l$. 
as ${\gamma}_l= {\gamma}_{\psi}\pi \ln (\lambda /{\xi}_s)
$.\cite{kk}   
Hence the weak chirality regime, eq.(\ref{ptd}) reduces to the 
stochastic equation for a single defect line, 
${\gamma }_l \frac{d {\bf R}_l }{dt} =
-\frac{\delta F}{\delta {\bf R}_l}+ {\bf G}_{l}(t)$.  
The nois ${\bf G}_{l}(t)$ satisfies  
$<G_l(t) G_l(t') >=\frac{2k_BT}{{\gamma}_{l}} \delta(t-t')$. 
Later we will show that the velocity $v_l=\frac{\eta \chi dE}{{\gamma}_l}$ is obtained 
from the experimental parameters 
$\eta $, $\chi$, $d$. 

At $k_0\sim k_{c2}$ 
($l_bl_d \sim {\xi}_s^2$) the screw dislocations cannot be treated 
as independent mobile entities. 
The static solutions of the linearized GL equation for $\psi$ 
in the superconductor case near ${\bf H}_{c2}$ can be approximated by 
eigenfunctions of the lowest Landau level. 
An analogous procedure can be taken for the TGBA case when $q_0/l_0 \gg 1$. \cite{rl}
Here we add the time dependence into the solution. 
\begin{eqnarray}
\lefteqn{
\psi ({\bf x}) = } \cr
& & {\sum}_s C_s \exp \left( i{\bf q}_{\perp}^s(x_s+U_s(t)) \cdot {\bf x} \right) \cr 
& &
\exp \left( -\frac{(k_0x-k_0(x_s+U_s(t)))^2 }{{\bar{\xi }}^2} \right)  \label{kc2psi}
\end{eqnarray} 
Here, $x_s$ is the position of $s$-th SmA slab, 
$U_s(t)$ denotes the displacement of the zero of the $s$-th
eigenfuntion in the lowest Landau level. 
${\bf q}_{\perp}^s(x_s+U_s(t))
=q_0(0, \sin(k_0(x_s+U_s(t)), \cos (k_0(x_s+U_s(t)))$, 
$\bar{\xi}^2=1/(q_0k_0)$.
The dissipation function $R$ per unit volume in this case 
$R= \frac{{\gamma}_{\psi}q_0k_0v_l^2\sqrt{\pi}}{l_bq_0}$. 
As we will discuss next, the velocity $v_l$ (now defined as the 
average velocites of the zeroes) is decided by combining $R$ and
the electric energy dissipation.


First we point out that a d.c. polarization current is directly
observable  
as a direct consequence of the motion of screw dislocations. 
This is in sharp contrast to dislocation-free feroelectrics, 
where only a.c. currents are allowed to flow. 
The origin of the charge carried by the screw dislocation 
stems from the chiral nature of the system. 
As can be seen from eq.(1), the deviation of the director vector from 
the layer normal gives rise to polarization. 
Around a screw dislocation 
this is obtained as 
${\bf P}=-\frac{\eta d}{2\pi {\lambda }_2 } 
{\cal K}_0 \left( \frac{\mid {\bf x}-{\bf R}_l \mid }{{\lambda }_2} \right) 
{\bf e}_r +\chi {\epsilon}_0{\bf E}$ per unit length,  
where ${\cal K}_0$ is the modified Bessel function, 
${\bf e}_r=(\cos \phi, \sin \phi, 0)$, 
and $\phi = {\tan }^{-1}\left( \frac{y-{\bf R}_{ly}}{x-{\bf R}_{lx}} \right)$. 
Hence at the surface of the core of a screw dislocation 
we have a local charge accumulation of 
$Q= \int d{\bf S} \nabla \cdot {\bf P} =\eta \chi d$. 
Thus near $k_{c1}$ we are able to relate the velocity 
to the polarization current by summing up the contribution 
$\partial_tQ$ and get 
\begin{equation}
{\bf j}_p =\frac{\eta \chi d v_l\hat{\bf x}}{l_dl_b}
=\eta \chi {v}_l \bar{k}_0 \hat{\bf x}_1. \label{jpkc1}
\end{equation}
Here, $\bar{k}_0$ is the spatialy averaged twist of the director, 
for which the relation $\bar{k}_0=\frac{d}{l_bl_d}$ for small $\alpha $ is used. 
Near $k_{c2}$ ($\sqrt{l_dl_b} \sim {\xi }_s$ ) 
using the solution of eq.~(\ref{kc2psi}) leads to 
the same form as in eq.~(\ref{jpkc1}). 
As is the case of superconductors 
$\bar{k}_0$ is expressed in the form of the expansion of 
$\frac{k_{c2}-k_0}{k_0}$ near $k_{c2}$. 
The parameter $\bar{\beta }$\cite{rl} (corresponding to 
Abrikosov's beta in superconductors) is related to the configuration of 
screw dislocation lattice. 
Differentiating the free energy with respect to  $K_2k_0$, we obtain   
$\bar{k}_0 = k_0- \frac{k_{c2}-k_0}{2 {\kappa}_2^2 \bar{\beta}}$
with ${\kappa}_2 = \frac{1}{Cq_0} \sqrt{\frac{gK_2}{2}}$. 
Demanding the dissipation 
due to the polarization current to coincide with $R$ which is induced by 
the defects dynamics, 
we are able to express $v_l$ as  
$v_l = \frac{\eta \chi dE l_bq_0 \bar{k}_0 }{{\gamma}_{\psi}k_0l_d\sqrt{\pi}}$ 
in terms of the material parameters 
$k_0$, $\eta$, $\chi$, and viscosity${\gamma}_{\psi}$. 

In summary we have investigated the basic electric
properties of the dislocation motion in the TGBA phase, 
with a special emphasis on the analogy with vortex dynamics in 
superconductors. 
In actual experimental situations, 
rubbing process to fix the direction of directors of 
molecules should not be made in order to make the predicted effects observable. 
The present work should be considered a starting point 
for further studies of the dynamics in the TGB system, in which 
the interplay between thermal fluctuation and defect dynamics 
is expected to give rise to various interesting physics. 


It is pleasure to acknowledge many helpful discussions with 
Professor T. Tokihiro at the early stage of this work.
A. T. acknowledge a fellowship from the Japan Atomic Energy Research Institute.
T. O. was supported in part by the Japan Society for the Promotion
of Science.

n


\begin{figure}
\caption{ 
The stress field near the grain boundary. 
${\bf J}_v$ is the nonvanishing curl stress field 
around screw dislocations with its separeation $l_d$.
${\bf J}_e^L$ and ${\bf J}_e^R$ is the electric field induced spatially uniform stress field of left SmA slab and right SmA slab, respectively.  
}
\label{fig1}
\end{figure}

\begin{figure}
\caption{ 
 The smectic layers near the grain boundary. 
$d$ is layer separation of SmA slabs. 
The angle between electric field induced stress field in left SmA slab ${\bf J}_e^L$ along the Sm layer and 
that of right SmA slab ${\bf J}_e^R$ is $2\pi \alpha $. 
}
\label{fig2}
\end{figure}

\begin{table}
\begin{tabular}{ll}
\hline
 superconductors &liquid crystals \cr
\hline
 ${\bf j}_s=-\frac{\delta F}{\delta {\bf A}} $:supercurrent 
&${\bf j} =-\frac{\delta F}{\delta {\bf n}} $:stress  \cr
 Lorentz force& Peach-Koehler force \cr
 vortex lattice flow & grain boundary flow \cr
& due to the electric field \cr
 Nernst effect & grain boundary flow \cr
& due to the temperature gradient \cr
flux flow resistivity 
& conductivity due to grain boundary flow \cr
\hline 
\end{tabular}
\caption[ The analogy between superconductors and liquid crystals ]
{ Dynamical aspects of the analogy between superconductors and liquid crystals.} 

\label{table1}
\end{table}

\end{document}